\documentclass[]{iopart}

\usepackage[backref,breaklinks,colorlinks,citecolor=blue]{hyperref}  
\usepackage[all]{hypcap}

\newcommand{\msun}{\ensuremath{M_\odot}}

\usepackage{amssymb}
\usepackage{graphicx}
\usepackage{color}
\usepackage{natbib}
\usepackage{xspace}
\usepackage{nicefrac}
\usepackage[dvipsnames]{xcolor}
\newcommand\myshade{85}
\usepackage[T1]{fontenc}
\usepackage{lmodern}

\colorlet{mycitecolor}{Turquoise}
\colorlet{mylinkcolor}{Turquoise}

\hypersetup{
 citecolor = mycitecolor!\myshade!black,
 linkcolor = mylinkcolor!\myshade!black,
 colorlinks = true,
}

\def\be{\begin{equation}}
\def\ee{\end{equation}}
\def\ba{\begin{eqnarray}}
\def\ea{\end{eqnarray}}

\begin{document}

\title{The astrophysical science case for a decihertz gravitational-wave detector}

\def\addBham{Institute of Gravitational Wave Astronomy and School of Physics and Astronomy, University of Birmingham, Edgbaston, Birmingham B15 2TT, United Kingdom}

\author{Ilya Mandel$\footnote[1]{email:imandel@star.sr.bham.ac.uk}$, Alberto Sesana, Alberto Vecchio}
\address{\addBham}

\begin{abstract}

We discuss the astrophysical science case for a decihertz gravitational-wave mission.  We focus on unique opportunities for scientific discovery in this frequency range, including probes of type IA supernova progenitors, mergers in the presence of third bodies, intermediate mass black holes, seeds of massive black holes, improved sky localization, and tracking the population of merging compact binaries.

\end{abstract}

\maketitle

\section{Introduction}

The recent detections of gravitational waves from merging binary black holes \citep{GW150914} and binary neutron stars \citep{GW170817} by the ground-based advanced LIGO \citep{AdvLIGO} and Virgo \citep{AdvVirgo} gravitational-wave detectors have stimulated interest in the full spectrum of gravitational-wave astronomy.  Pulsar timing arrays are actively searching for gravitational waves in the nanohertz frequency band \citep{PTA,IPTA}; a space-born LISA detector, sensitive in the milihertz band, is scheduled to be launched in the 2030s \citep{LISA,LISA:2017}; and there are ongoing investigations into a future ground-based detector with improved low-frequency sensitivity reaching down to a few hertz, e.g., the Einstein Telescope \citep{ET}.   In this paper, we make the astrophysical case for a detector that would slot in between the LISA band and the Einstein Telescope band, with peak sensitivity around 1 decihertz.  This science case partly overlaps the cases already made for terrestrial detectors such as the Einstein Telescope \citep{ET:2012} and the MANGO detector \citep{Harms:2013}, as well as the proposed space missions DECIGO \citep{DECIGO}, ALIA and BBO \citep{TakahashiNakamura:2003,CrowderCornish:2005}.  

Here, we focus on the key science questions that may not be answered by the either ground-based detectors sensitive above 1 Hz or millihertz space detectors, but could be addressed by decihertz instruments.  We do not consider any specific instruments with associated noise spectra, although a broad range of recent proposals, from the TianQin space detector \citep{TianQin} to atom interferometers \citep[e.g.,][]{Graham:2013}, could be sensitive in the band of interest.  Instead, we focus on the main scientific challenges, and where appropriate discuss the sensitivities necessary to address these.  

In particular, we highlight the promise of decihertz detectors to pinpoint the progenitors of type IA supernovae; search for dynamical signatures of the merger envrionment; explore intermediate mass black holes; localize compact binaries on the sky; explore the evolutionary history of stellar-mass compact-object binaries;  and investigate the light seeds of massive black holes.

\section{Type IA supernova progenitors}

Do type IA supernovae come from the merger of two white dwarfs (the double degenerate channel) or from accretion onto a white dwarf from a main sequence or giant companion (the single degenerate channel) \citep[e.g.,][]{Livio:2000,PodsiadlowskiHan:2004,Nielsen:2014}?  This has been a topic of active debate with differing interpretations of the observational evidence in the literature \citep[e.g.,][]{GilfanovBogdan:2010,Hayden:2010,Mennekens:2010,Nugent:2011,GonzalezHernandez:2012,Howell:2011}.  

Gravitational-wave observations in the decihertz band could help resolve this question.  Joint observations of GW emission and a supernova would indicate the double degenerate channel, while the absence of a gravitational-wave signal preceding a nearby type IA would indicate the single degenerate channel, as the stellar companion would have been disrupted at lower frequencies.  

The gravitational-wave frequency for a circular binary with total mass $M$ and orbital separation $a$ is given by
\begin{equation}
f_\textrm{GW} \approx 0.1\, \textrm{Hz} \left(\frac{M}{2M_\odot}\right)^{1/2}  \left(\frac{a}{0.02 R_\odot}\right)^{-3/2}.
\end{equation}
Depending on the companion mass, a double WD binary could survive until it reaches an orbital radius $\sim 0.02 R_\odot$ (see \citet{Dan:2011} for somewhat lower numerical estimates of the maximum gravitational-wave frequency).  However, if the white dwarf's companion is a main sequence star or a giant, the companion would be disrupted at much larger separations.  Therefore, the presence of gravitational waves in the decihertz band would be a tell-tale sign for the double degenerate channel.  (An explosion could be delayed by as much as $10^5$ years following the merger in the double-degenerate channel \citep{Yoon:2007}, in which case one would not expect a correlation between gravitational waves and a type IA supernova even if this channel is operating, but more recent work suggests that prompt post-merger explosions are likely \citep{Pakmor:2012}.  Meanwhile, if the double-degenerate channel proceeds via head-on white dwarf collisions in triples \citep{Kushnir:2013}, there may not be a strong associated gravitational-wave signature.)

The rate of type IA supernovae is roughly 1 per century per Milky Way equivalent galaxy \citep{Cappellaro:1999}, while the space density of such galaxies is $\sim 0.01$ Mpc$^{-3}$  \citep{LIGOS3S4Galaxies}.  Therefore, to have a realistic chance of observing a t least one type IA supernova per year, $\sim 10^4$ Mpc$^3$ must be surveyed -- roughly the volume out to the Virgo cluster.  (In fact, this would yield a slightly greater rate because of the local over-density of galaxies \citep{LIGOS3S4Galaxies}, which would compensate for the possible non-detection of some nearby supernovae due to unfavourable sky locations, etc.)  Hence the gravitational-wave detector should also be sensitive out to $\sim 20$ Mpc for such signals on average -- or out to $\sim 50$ Mpc for optimally located and oriented events \citep{Finn:1996}.

The amplitude of the frequency-domain gravitational-wave signal from a binary inspiral viewed face on is \citep{Ajith:2008}
\begin{equation}
|\tilde{h}(f)| = \sqrt{\frac{5}{24 \pi^{4/3}}} M_c^{5/6} f^{-7/6} r^{-1},
\end{equation}
where the chirp mass $M_c = 2^{-1.2} M_\odot$ for an equal-mass binary, $r$ is the distance to the source, and dimensionless units $G=c=1$ are assumed.  The signal-to-noise ratio $\rho$ for a detector with one-sided noise power spectral density $S_n(f)$ is given by
\begin{equation} 
\rho^2 = 4 \int_0^\infty \frac{|\tilde{h}(f)|^2}{S_n(f)} df.
\ee

We can now use these expressions to check whether a given noise spectrum would be sufficient to allow a double white dwarf binary with $M_c \sim M_\odot$ to be detected out to $r_\textrm{max} \sim 50$ Mpc at optimal location and orientation.  For example, assuming that the detector has a flat noise power spectral density $S_n(f) = S$ between $f_\textrm{min}$ and $f_\textrm{max}$ and limited sensitivity elsewhere, the sensitivity requirement on $S$ for a detection threshold $\rho_\textrm{min}$ is
\begin{equation}\label{SNR}
S \lesssim 2\times 10^{-42} \textrm{Hz}^{-1} \left[ \left(\frac{f_\textrm{min}}{0.1 \textrm{Hz}}\right)^{-4/3} - \left(\frac{f_\textrm{max}}{0.1 
\textrm{Hz}}\right)^{-4/3} \right]
	\left(\frac{M_c}{M_\odot}\right)^{5/3} \left(\frac{r_\textrm{max}}{50 \textrm{Mpc}}\right)^{-2} \left(\frac{\rho_\textrm{min}}{8}\right)^{-2};
\end{equation}
for $f_\textrm{min}=0.1$ Hz and $f_\textrm{max}=0.2$ Hz, we find that $S \lesssim 10^{-42}$ Hz$^{-1}$ if the detection threshold is $\rho_\textrm{min} \geq 8$.    

\begin{figure*}[htb]
\centering
\includegraphics[width=1.0\linewidth,angle=0]{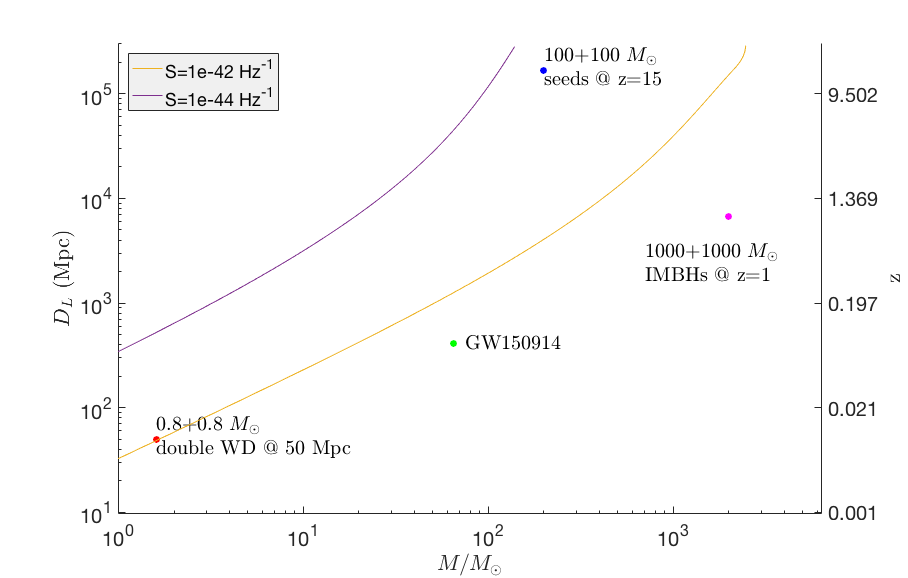}
\caption{Maximum reach of an instrument with the specified noise power spectral density, $S=10^{-42}$ ($S=10^{-44}$) Hz$^{-1}$ for the bottom (top) curve, to a binary with equal component masses and a total mass as specified on the abscissa, for an integration band from $0.1$ to $0.2$ Hz and a signal-to-noise ratio threshold of $8$.   The left ordinate shows the luminosity distance; the right ordinate the corresponding redshift.  A few selected sources are labeled for reference, including GW150914 at its observed distance \citep{GW150914}.}
\label{fig_max}
\end{figure*}

Equation (\ref{SNR}) can be inverted to obtain the distance reach $r_\textrm{max}$ as a function of $S$.  This reach is shown in Figure \ref{fig_max}.  At cosmological scales, $r_\textrm{max}$ in Equations (\ref{SNR}) and (\ref{SNRf}) becomes the luminosity distance, while the masses are the redshifted masses, $M \to M(1+z)$, where $z$ is the redshift.

The timescale until a gravitationally-wave driven merger for a circular binary with a current gravitational-wave frequency $f_\textrm{GW}$ is \citep{Peters:1964}
\begin{equation}\label{time}
\tau_\textrm{GW} \approx 5\ \textrm{yr} \left(\frac{M_c}{M_\odot}\right)^{-5/3} \left(\frac{f_\textrm{GW}}{0.1 \textrm{Hz}}\right)^{-8/3}.
\end{equation}
Thus, a binary with a chirp mass of $1 M_\odot$ would need almost 5 years to evolve from a gravitational-wave frequency of $0.1$ Hz to a gravitational-wave frequency of $0.2$ Hz.  At even lower frequencies, when the evolutionary timescale $\tau_\textrm{GW}$ is much longer than the observing duration $T_\textrm{obs}$, so that the signal can be considered as roughly monochromatic at frequency $f_\textrm{GW}$, the sensitivity requirement given in Equation (\ref{SNR}) is modified to 
\begin{equation}\label{SNRf}
S  \lesssim  5 \times 10^{-43} \textrm{Hz}^{-1} \left(\frac{M_c}{M_\odot}\right)^{10/3} \left(\frac{f_\textrm{GW}}{0.05 \textrm{Hz}}\right)^{4/3} \left(\frac{T_\textrm{obs}}{5\ \textrm{yr}}\right)  \left(\frac{r_\textrm{max}}{50 \textrm{Mpc}}\right)^{-2} \left(\frac{\rho_\textrm{min}}{8}\right)^{-2}.
\end{equation}

\section{Mergers in the presence of third bodies}

Low-frequency, long-duration observations are potentially sensitive to astrophysical perturbations to gravitational-wave driven binary evolution, such as Doppler shifting of the gravitational-wave signature due to the orbital motion of the inspiraling binary relative to a third companion in the system.  The Doppler shift is given by
\begin{equation}
\delta f \approx 10^{-5} \textrm{Hz} \frac{f_\textrm{GW}}{0.1 \textrm{Hz}} \frac{m}{M_\odot} \left(\frac{M_\textrm{bin}}{M_\odot}\right)^{-0.5}  \left(\frac{a}{\textrm{AU}}\right)^{-0.5},
\end{equation}
where $M_\textrm{bin}$ is the mass of the merging compact binary emitting gravitational waves at frequency $f_\textrm{GW}$ and $m \ll M_\textrm{bin}$ is the mass of the tertiary companion at a separation $a$ from the binary.  Fluctuations $\delta f \gtrsim 1/(\rho T_\textrm{obs})$ should be detectable as long as the observation time $T_\textrm{obs}$ is larger than the outer orbital period $2\pi a^{3/2} (G M)^{-1/2}$.  The readily inferred presence of a third companion could indicate the importance of the Lidov-Kozai mechanism \citep{Lidov:1962,Kozai:1962} in driving binaries to merger.  

Conversely, if the binary is merging within the sphere of influence of a massive black hole --- the merger of a stellar-mass binary may be assisted by the accretion disk in an active galactic nucleus \citep{Bartos:2016,Stone:2016} --- the orbital period around the massive black hole is typically much longer than the observation duration.  A constant Doppler shift is degenerate with a cosmological redshift or a change in the mass of the binary.  However, the orbital acceleration of the binary around the massive black hole of mass $M_\textrm{MBH} \gg M_\textrm{bin}$ will be detectable when the accumulated acceleration-induced phase shift to the gravitational-wave signal exceeds $\sim 1/\rho$, 
\begin{equation}
\frac{1}{2}\frac{G M_\textrm{BH}}{a^2}{T_\textrm{obs}^2}\frac{f_\textrm{GW}}{c} \gtrsim \frac{1}{\rho},
\end{equation}
i.e., when
\begin{equation}
f_\textrm{GW} \gtrsim 0.02\, \textrm{Hz} \left(\frac{M_\textrm{MBH}}{10^6\ M_\odot}\right)^{-1}\, \left(\frac{a}{\textrm{pc}}\right)^{2} \left(\frac{T_\textrm{obs}}{5 \textrm{yr}}\right)^{-2} \left(\frac{\rho}{8}\right)^{-1}
\end{equation}
for a suitable binary orientation relative to the line of sight.  Thus, a decihertz gravitational waves from a double neutron star or double white dwarf inspiraling within a massive black hole's sphere of influence will carry the signature of its environment.  On the other hand, the merger timescale from $f_\textrm{GW}=0.1$ Hz for a binary black hole is much shorter than 5 years (see Equation (\ref{time})).  Therefore, to detect the imprint of the massive black hole on the gravitational waves from a merging stellar-mass black hole binary, either the detector sensitivity would need to extend down to $\sim 0.01$ Hz, or the merger would need to happen within $\sim 1000$ AU of the massive black hole.  

The density in the center of the most massive core-collapsed globular clusters is comparable to the mass concentration within the sphere of influence of a massive black hole; therefore, gravitational-wave signatures of decihertz binaries in globular clusters may also carry an imprint of their environment.

\section{Intermediate mass black holes}

A decihertz mission could be the optimal tool for searching for $\sim 1000 M_\odot$ intermediate mass black holes (IMBHs).  Black holes in this mass range are notoriously challenging to convincingly find.   Their small sphere of influence means that only a handful of nearby objects show unambiguous dynamical impact of the  IMBH \citep[e.g.,][]{MillerColbert:2004,Kiziltan:2017,Freire:2017}. Meanwhile, possible ultra-luminous X-ray binaries could be interpreted as either IMBHs \citep[e.g.,][]{Pasham:2014} or super-Eddington accretors \citep{Bachetti:2014}.   Gravitational-wave observations of either inspirals of stellar-mass compact objects into IMBHs, or mergers of two IMBHs, could thus provide the first convincing evidence of their existence.   

The gravitational-wave frequency from an innermost stable circular orbit around a black hole of mass $M$ is
\begin{equation}
  f_\textrm{ISCO} \approx 4.3  \textrm{Hz} \left(\frac{M}{1000\ M_\odot}\right)^{-1},
  \label{eq:fisco}
\end{equation}
placing such massive black holes outside the range of ground-based detectors insensitive below a few Hz \citep{Gair:2009ETrev,Belczynski:2014VMS}, but into the range of decihertz detectors.     

If black holes in this mass range naturally reside in globular clusters, intermediate-mass ratio inspirals should be generic \citep[e.g.,][]{Haster:2016}, and the mass of the black hole could be confirmed through the associated gravitational-wave signature \citep{Haster:2015IMRI}.  The local space density of globular clusters is a few per Mpc$^3$, and an upper limit on the merger rate can be estimated by assuming that the intermediate mass black hole builds up its mass through minor mergers over the $\sim 10^{10}$ yr cluster lifetime \citep{Mandel:2008}.  Thus, if a few percent of all globular clusters host a $\sim 1000 M_\odot$ black hole, an intermediate mass ratio coalescence of such an IMBH and a $\sim 10 M_\odot$ companion may occur at a rate of up to one merger per Gpc$^3$ per year.  The detector described above would be sensitive to these coalescences at Gpc-scale distances, and could therefore realistically detect such inspirals and confirm the existence of intermediate mass black holes in this mass range.  Such confirmation could also come from mergers of IMBH binaries \citep{Amaro:2006imbh}.

Observations of coalescences involving IMBHs would enable exploration of globular cluster dynamics.  These coalescences could also provide electromagnetic counterparts if the inspiraling compact object is a white dwarf rather than a neutron star or black hole \citep{Sesana:2008}.  An inspiral of a white dwarf into an intermediate mass black hole in the $10^4 M_\odot$ range could be detectable to $z \gtrsim 0.5$ for a detector with noise power spectral density $S \sim 10^{-43}$ Hz$^{-1}$ between $0.1$ and $0.2$ Hz; \citep{Sesana:2008} argue that at least a few such inspirals should happen in this range per year.  Such white dwarf tidal disruptions have been proposed as a possible source of a recently observed population of faint X-ray transients \cite{Bauer:2017}.



\section{Massive black hole formation}
Decihertz detectors could look for gravitational waves from light seeds of today's massive black holes (MBHs). MBHs inhabit the center of essentially all massive galaxies in the nearby Universe \citep{Kormendy:1995,Magorrian:1998}, and their masses correlate with the properties of the galaxy host, pointing toward  MBH-host co-evolution \citep[see][and references therein]{KormendyHo:2013}. This implies that following galaxy mergers, MBHs form MBH binaries \citep{Begelman:1980}, which are expected to be loud sources of GWs. The merger rate of such binaries strongly depends on the early occupation fraction of the first MBH 'seeds' and on their masses \citep{VHM,Sesana:2011}. In particular, different scenarios for forming the first BH seeds have been proposed in the literature \citep[see][for a review]{Volonteri:2010}. Seeds forming from the direct collapse of protogalactic disks in the mass range $10^4\msun-10^5\msun$ are ideal targets for the LISA mission. One the other hand, the decihertz band is the ideal window to catch potentially lower $\approx 100\msun$ seeds, left behind by the first generation of stars (Pop III).  At $z \sim 10$, the observed merger frequency of those binaries is approximately $1$Hz, according to Equation (\ref{eq:fisco}); therefore, some of the seeds, merging later / at lower redshifts, may even be observable with the Einstein Telescope \citep{Sesana:2009ET,Gair:2009ET}.    

Decihertz detectors should be sensitive to  such binaries, making them invaluable probes of structure formation. As shown in Figure \ref{fig_seed}, a detector with noise power spectral density $10^{-42}$ Hz$^{-1}$ between $0.1$ and $0.2$ Hz would be sensitive to mergers of two optimally oriented $2000 M_\odot$ seeds out to $z \sim 20$, whereas a spectral density of  $\sim 10^{-44}$ Hz$^{-1}$ in the same frequency range would be sufficient to cover the relevant mass range down to seeds of $100 M_\odot$, thus directly probing the very first seed BH mergers.
  
\begin{figure*}
\centering
\begin{tabular}{cc}
  \includegraphics[width=0.5\linewidth,clip=true,angle=0]{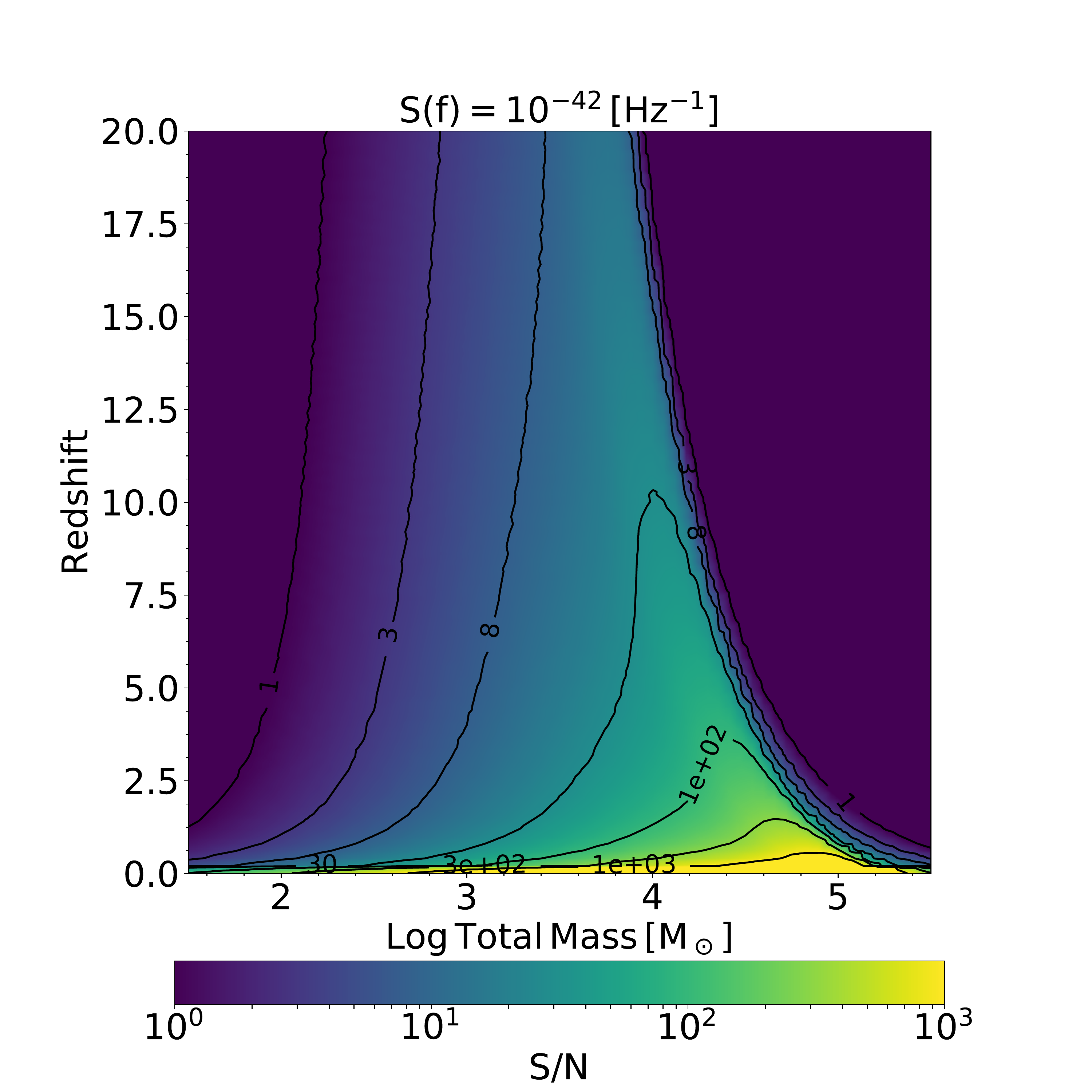}&
  \includegraphics[width=0.5\linewidth,clip=true,angle=0]{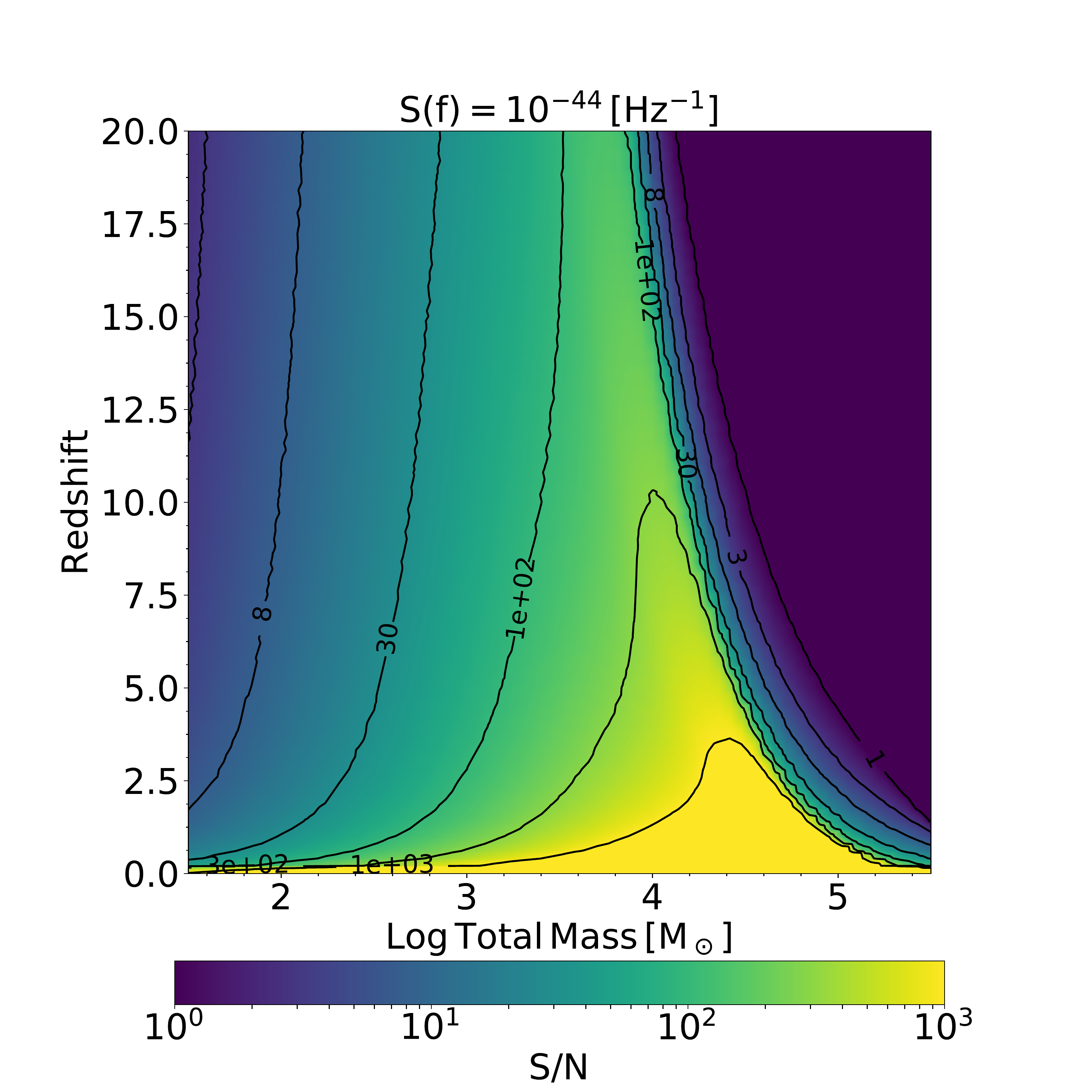}\\
\end{tabular}
\caption{Signal-to-noise ratio contour plots for optimally oriented equal mass black hole binaries with total mass indicated on the abscissa merging at a redshift indicated on the ordinate. The two plots depict the capabilities of detectors with $S = 10^{-42}$ Hz$^{-1}$ (left panel) and $S = 10^{-44}$ Hz$^{-1}$ (right panel) between $f_\textrm{min}=0.1$ Hz and $f_\textrm{max}=0.2$ Hz .}
\label{fig_seed}
\end{figure*}
  
\section{Sky localization}

A double neutron star emitting at a gravitational-wave frequency of $0.1$ Hz will only merge in several years according to Eq.~(\ref{time}).  A decihertz detector will thus complete several orbits around the Sun while the source is in band.  For sky localization purposes, such a detector effectively behaves as a set of detectors with a baseline of order the size of the orbit.  

The timing accuracy scales inversely with the detector bandwidth $f_\textrm{bandwidth}$ and inversely with the signal-to-noise ratio \citep{Fairhurst:2009,Grover:2013}; for a decihertz detector, $\sim 1$ s timing can be expected.  The sky localization accuracy can then be estimated as the timing accuracy divided by the light travel time across the detector baseline:
\begin{equation}
\sigma_\theta \sim 0.0025 \frac{0.1\textrm{Hz}}{f_\textrm{bandwidth}} \frac{8}{\rho} \frac{\textrm{AU}}{\textrm{baseline}}.
\end{equation}
(This approximation is equivalent to considering the impact of Doppler shifting of the signal by the motion of the detector.)

For a multi-year source, a baseline of 2 AU or 1000 light seconds yields a sky localization accuracy of $\sim 0.001$ radians or a few arcminutes -- though the accuracy of the localization will depend on the location of the source relative to the detector's orbital plane.   This would make it possible to accurately point smaller field-of-view, sensitive telescopes for electromagnetic follow-up.  It would even allow for host galaxy identification for nearby, $r \lesssim 500$ Mpc, sources, allowing host environments to be explored even in the absence of a confirmed counterpart.    

A heavy stellar-mass black hole binary with $\sim 30 M_\odot$ components will merge in about a week from a gravitational-wave frequency of $0.1$ Hz.  Its baseline will be much shorter -- only 10 light seconds -- and the arc is almost a straight line, with minimal perpendicular displacement to provide orthogonal directional sensitivity; therefore, accurate localization and host identification would remain challenging.

\section{Evolutionary history of compact object binaries}

Together with other gravitational-wave instruments -- space-born LISA, and ground-based advanced LIGO \citep{AdvLIGO} and Virgo \citep{AdvVirgo} detectors and their successors such as the proposed Einstein Telescope \citep{ET} -- a decihertz detector can ensure that the full frequency spectrum is covered for stellar-mass binary black holes and neutron stars.  

Multi-frequency observations can improve the accuracy with which source parameters can be measured.     Tracking an individual source across a range of frequencies could yield both information that is most readily accessible at higher frequencies and at lower ones.  For example, tidal effects for neutron stars or the total mass and final spin from the ringdown of a post-merger black hole would be measured with high-frequency observations.   On the other hand, as discussed above, sky localization can be significantly enhanced with lower-frequency observations.   

Of particular interest are measurements of spin magnitude and misalignment angle distributions, which could carry information about formation scenarios \citep[e.g.,][]{Stevenson:2017spin,Zevin:2017,Farr:2017} and the mass ratios, which would help constrain masses and test for the existence of a mass gap between neutron stars and black holes \citep{Belczynski:2012,Mandel:2015,Littenberg:2015}.  The mass ratio and spin-orbit coupling come into the waveform at higher orders in the orbital velocity \citep{PoissonWill:1995} and may therefore be better constrained at higher frequencies.   However, the presence of $\sim 10^5$ (heavy stellar-mass black hole binaries) to $\sim 10^7$ (double neutron stars) cycles in the decihertz band could, in fact, allow for more precise constraints.  Specific detector performance would need to be considered for a quantitative assessment of parameter inference with decihertz observations.

In addition to individual sources, it may be possible to track changes in source populations as they evolve between different frequency bands.  For example, residual eccentricity, which would strongly indicate dynamical formation \citep[e.g.,][]{GW150914:astro}, may only be observable at lower frequencies, as binaries will circularize through gravitational-wave emission by the time they reach the frequency band of classical ground-based detectors.  On the other hand, very eccentric sources at low frequencies could be more difficult to detect \citep{Chen:2017}, and their emergence at higher frequencies would indicate high birth eccentricity.

\section{Stochastic background}

The stochastic background from a superposition of gravitational waves emitted by multiple individually unresolvable binary inspirals in this frequency band should be a known power law in frequency \citep{GW150914:stoch}.   The amplitude of this power law will be sensitive to mergers at a higher redshift than the individually resolvable source population, but will still add only a single number to the information gained from that population.   However, observations at these and higher frequencies may make it possible to remove the astrophysical background.  This would make a potential gravitational-wave background of cosmological original \citep[e.g.][]{Mandic:2012, Lasky:2016} accessible to observation \citep{Callister:2016}.

\section{Discussion}

We have outlined the exciting astrophysical potential of a decihertz detector.  We estimated the sensitivity required to achieve several key science goals that can best be addressed in this frequency band, including investigating the progenitors of type IA supernovae, measuring the dynamics in the merger environment, searching for intermediate-mass black holes and light seeds of today's massive black holes, and exploring the evolutionary history of compact object binaries. 

As always, exploring a new spectral band opens the potential for unexpected discoveries.  For instance, predicted but elusive gravitational waves from cosmic string cusps may be detectable in this band.

Decihertz observations may also enable more precise tests of general relativity, or at least tests in a different regime of velocities and tidal field strengths \citep[e.g.,][]{Chamberlain:2017}.  For example, the mass quadrupole moment of the compact bodies could be measured and compared with the value predicted from mass and spin in the Kerr metric \citep{Brown:2007,Rodriguez:2012}.

Our estimates will need to be followed up for specific proposed detector noise power spectra in order to quantitatively evaluate the prospects discussed above.  In particular, rigorous estimates of parameter inference accuracy combined with realistic astrophysical models will be required to appraise the resolving power of intermediate-frequency GW detectors.  

\section*{Acknowledgments}
IM thanks Will Farr and Stephen Justham for discussions and Christopher Berry for comments on the manuscript.  IM and AV is partially supported by STFC.  AS holds a Royal Society University Research Fellowship.

\bibliography{Mandel}

\begin{thebibliography}{78}
\expandafter\ifx\csname natexlab\endcsname\relax\def\natexlab#1{#1}\fi

\bibitem[{{Aasi} {et~al.}(2015){Aasi}, {Abbott}, {Abbott}, {Abbott},
  {Abernathy}, {Ackley}, {Adams}, {Adams}, {Addesso}, \& et~al.}]{AdvLIGO}
{Aasi}, J. {et~al.} 2015, Classical and Quantum Gravity, 32, 074001, 1411.4547

\bibitem[{{Abbott} {et~al.}(2016{\natexlab{a}}){Abbott}, {Abbott}, {Abbott},
  {Abernathy}, {Acernese}, {Ackley}, {Adams}, {Adams}, {Addesso}, {Adhikari},
  \& et~al.}]{GW150914:astro}
{Abbott}, B.~P. {et~al.} 2016{\natexlab{a}}, \apjl, 818, L22, 1602.03846

\bibitem[{{Abbott} {et~al.}(2016{\natexlab{b}}){Abbott}, {Abbott}, {Abbott},
  {Abernathy}, {Acernese}, {Ackley}, {Adams}, {Adams}, {Addesso}, {Adhikari},
  \& et~al.}]{GW150914:stoch}
------. 2016{\natexlab{b}}, Physical Review Letters, 116, 131102, 1602.03847

\bibitem[{{Abbott} {et~al.}(2016{\natexlab{c}}){Abbott}, {Abbott}, {Abbott},
  {Abernathy}, {Acernese}, {Ackley}, {Adams}, {Adams}, {Addesso}, {Adhikari},
  \& et~al.}]{GW150914}
------. 2016{\natexlab{c}}, Physical Review Letters, 116, 061102, 1602.03837

\bibitem[{Acernese {et~al.}(2015)}]{AdvVirgo}
Acernese, F., {et~al.} 2015, Class. Quant. Grav., 32, 024001, 1408.3978

\bibitem[{{Ajith} {et~al.}(2008){Ajith}, {Babak}, {Chen}, {Hewitson},
  {Krishnan}, {Sintes}, {Whelan}, {Br{\"u}gmann}, {Diener}, {Dorband},
  {Gonzalez}, {Hannam}, {Husa}, {Pollney}, {Rezzolla}, {Santamar{\'{\i}}a},
  {Sperhake}, \& {Thornburg}}]{Ajith:2008}
{Ajith}, P. {et~al.} 2008, \prd, 77, 104017, 0710.2335

\bibitem[{{Amaro-Seoane} {et~al.}(2017){Amaro-Seoane}, {Audley}, {Babak},
  {Baker}, {Barausse}, {Bender}, {Berti}, {Binetruy}, {Born}, {Bortoluzzi},
  {Camp}, {Caprini}, {Cardoso}, {Colpi}, {Conklin}, {Cornish}, {Cutler},
  {Danzmann}, {Dolesi}, {Ferraioli}, {Ferroni}, {Fitzsimons}, {Gair}, {Gesa
  Bote}, {Giardini}, {Gibert}, {Grimani}, {Halloin}, {Heinzel}, {Hertog},
  {Hewitson}, {Holley-Bockelmann}, {Hollington}, {Hueller}, {Inchauspe},
  {Jetzer}, {Karnesis}, {Killow}, {Klein}, {Klipstein}, {Korsakova}, {Larson},
  {Livas}, {Lloro}, {Man}, {Mance}, {Martino}, {Mateos}, {McKenzie},
  {McWilliams}, {Miller}, {Mueller}, {Nardini}, {Nelemans}, {Nofrarias},
  {Petiteau}, {Pivato}, {Plagnol}, {Porter}, {Reiche}, {Robertson},
  {Robertson}, {Rossi}, {Russano}, {Schutz}, {Sesana}, {Shoemaker}, {Slutsky},
  {Sopuerta}, {Sumner}, {Tamanini}, {Thorpe}, {Troebs}, {Vallisneri},
  {Vecchio}, {Vetrugno}, {Vitale}, {Volonteri}, {Wanner}, {Ward}, {Wass},
  {Weber}, {Ziemer}, \& {Zweifel}}]{LISA:2017}
{Amaro-Seoane}, P. {et~al.} 2017, ArXiv e-prints, 1702.00786

\bibitem[{{Amaro-Seoane} \& {Freitag}(2006)}]{Amaro:2006imbh}
{Amaro-Seoane}, P., \& {Freitag}, M. 2006, \apjl, 653, L53,
  arXiv:astro-ph/0610478

\bibitem[{{Ando} {et~al.}(2010){Ando}, {Kawamura}, {Seto}, {Sato}, {Nakamura},
  {Tsubono}, {Takashima}, {Funaki}, {Numata}, {Kanda}, {Tanaka}, {Ioka},
  {Agatsuma}, {Aoyanagi}, {Arai}, {Araya}, {Asada}, {Aso}, {Chiba},
  {Ebisuzaki}, {Ejiri}, {Enoki}, {Eriguchi}, {Fujimoto}, {Fujita}, {Fukushima},
  {Futamase}, {Harada}, {Hashimoto}, {Hayama}, {Hikida}, {Himemoto},
  {Hirabayashi}, {Hiramatsu}, {Hong}, {Horisawa}, {Hosokawa}, {Ichiki},
  {Ikegami}, {Inoue}, {Ishidoshiro}, {Ishihara}, {Ishikawa}, {Ishizaki}, {Ito},
  {Itoh}, {Izumi}, {Kawano}, {Kawashima}, {Kawazoe}, {Kishimoto}, {Kiuchi},
  {Kobayashi}, {Kohri}, {Koizumi}, {Kojima}, {Kokeyama}, {Kokuyama}, {Kotake},
  {Kozai}, {Kunimori}, {Kuninaka}, {Kuroda}, {Maeda}, {Matsuhara}, {Mino},
  {Miyakawa}, {Miyamoto}, {Miyoki}, {Morimoto}, {Morisawa}, {Moriwaki},
  {Mukohyama}, {Musha}, {Nagano}, {Naito}, {Nakamura}, {Nakamura}, {Nakano},
  {Nakao}, {Nakasuka}, {Nakayama}, {Nakazawa}, {Nishida}, {Nishiyama},
  {Nishizawa}, {Niwa}, {Noumi}, {Obuchi}, {Ohashi}, {Ohishi}, {Ohkawa},
  {Okada}, {Okada}, {Oohara}, {Sago}, {Saijo}, {Saito}, {Sakagami}, {Sakai},
  {Sakata}, {Sasaki}, {Sato}, {Shibata}, {Shinkai}, {Somiya}, {Sotani},
  {Sugiyama}, {Suwa}, {Suzuki}, {Tagoshi}, {Takahashi}, {Takahashi},
  {Takahashi}, {Takahashi}, {Takahashi}, {Takahashi}, {Takahashi}, {Akiteru},
  {Takano}, {Taniguchi}, {Taruya}, {Tashiro}, {Torii}, {Toyoshima},
  {Tsujikawa}, {Tsunesada}, {Ueda}, {Ueda}, {Utashima}, {Wakabayashi}, {Yagi},
  {Yamakawa}, {Yamamoto}, {Yamazaki}, {Yokoyama}, {Yoo}, {Yoshida}, {Yoshino},
  \& {Sun}}]{DECIGO}
{Ando}, M. {et~al.} 2010, Classical and Quantum Gravity, 27, 084010

\bibitem[{{Bachetti} {et~al.}(2014){Bachetti}, {Harrison}, {Walton},
  {Grefenstette}, {Chakrabarty}, {F{\"u}rst}, {Barret}, {Beloborodov}, {Boggs},
  {Christensen}, {Craig}, {Fabian}, {Hailey}, {Hornschemeier}, {Kaspi},
  {Kulkarni}, {Maccarone}, {Miller}, {Rana}, {Stern}, {Tendulkar}, {Tomsick},
  {Webb}, \& {Zhang}}]{Bachetti:2014}
{Bachetti}, M. {et~al.} 2014, \nat, 514, 202, 1410.3590

\bibitem[{{Bartos} {et~al.}(2016){Bartos}, {Kocsis}, {Haiman}, \&
  {M{\'a}rka}}]{Bartos:2016}
{Bartos}, I., {Kocsis}, B., {Haiman}, Z., \& {M{\'a}rka}, S. 2016, ArXiv
  e-prints, 1602.03831

\bibitem[{{Bauer} {et~al.}(2017){Bauer}, {Treister}, {Schawinski}, {Schulze},
  {Luo}, {Alexander}, {Brandt}, {Comastri}, {Forster}, {Gilli}, {Kann},
  {Maeda}, {Nomoto}, {Paolillo}, {Ranalli}, {Schneider}, {Shemmer}, {Tanaka},
  {Tolstov}, {Tominaga}, {Tozzi}, {Vignali}, {Wang}, {Xue}, \&
  {Yang}}]{Bauer:2017}
{Bauer}, F.~E. {et~al.} 2017, \mnras, 467, 4841, 1702.04422

\bibitem[{{Begelman} {et~al.}(1980){Begelman}, {Blandford}, \&
  {Rees}}]{Begelman:1980}
{Begelman}, M.~C., {Blandford}, R.~D., \& {Rees}, M.~J. 1980, \nat, 287, 307

\bibitem[{{Belczynski} {et~al.}(2014){Belczynski}, {Buonanno}, {Cantiello},
  {Fryer}, {Holz}, {Mandel}, {Miller}, \& {Walczak}}]{Belczynski:2014VMS}
{Belczynski}, K., {Buonanno}, A., {Cantiello}, M., {Fryer}, C.~L., {Holz},
  D.~E., {Mandel}, I., {Miller}, M.~C., \& {Walczak}, M. 2014, \apj, 789, 120,
  1403.0677

\bibitem[{{Belczynski} {et~al.}(2012){Belczynski}, {Wiktorowicz}, {Fryer},
  {Holz}, \& {Kalogera}}]{Belczynski:2012}
{Belczynski}, K., {Wiktorowicz}, G., {Fryer}, C.~L., {Holz}, D.~E., \&
  {Kalogera}, V. 2012, \apj, 757, 91, 1110.1635

\bibitem[{{Bender} {et~al.}(1998)}]{LISA}
{Bender}, P., {et~al.} 1998, LISA Pre-Phase A Report; Second Edition, Tech.
  Rep. MPQ233,
  http://list.caltech.edu/lib/exe/fetch.php?media=documents:early:prephasea.pdf

\bibitem[{{Brown} {et~al.}(2007){Brown}, {Brink}, {Fang}, {Gair}, {Li},
  {Lovelace}, {Mandel}, \& {Thorne}}]{Brown:2007}
{Brown}, D.~A., {Brink}, J., {Fang}, H., {Gair}, J.~R., {Li}, C., {Lovelace},
  G., {Mandel}, I., \& {Thorne}, K.~S. 2007, Physical Review Letters, 99,
  201102, gr-qc/0612060

\bibitem[{{Callister} {et~al.}(2016){Callister}, {Sammut}, {Qiu}, {Mandel}, \&
  {Thrane}}]{Callister:2016}
{Callister}, T., {Sammut}, L., {Qiu}, S., {Mandel}, I., \& {Thrane}, E. 2016,
  Physical Review X, 6, 031018, 1604.02513

\bibitem[{{Cappellaro} {et~al.}(1999){Cappellaro}, {Evans}, \&
  {Turatto}}]{Cappellaro:1999}
{Cappellaro}, E., {Evans}, R., \& {Turatto}, M. 1999, A\&A, 351, 459,
  arXiv:astro-ph/9904225

\bibitem[{{Chamberlain} \& {Yunes}(2017)}]{Chamberlain:2017}
{Chamberlain}, K., \& {Yunes}, N. 2017, ArXiv e-prints, 1704.08268

\bibitem[{{Chen} \& {Amaro-Seoane}(2017)}]{Chen:2017}
{Chen}, X., \& {Amaro-Seoane}, P. 2017, \apjl, 842, L2, 1702.08479

\bibitem[{{Crowder} \& {Cornish}(2005)}]{CrowderCornish:2005}
{Crowder}, J., \& {Cornish}, N.~J. 2005, \prd, 72, 083005, gr-qc/0506015

\bibitem[{{Dan} {et~al.}(2011){Dan}, {Rosswog}, {Guillochon}, \&
  {Ramirez-Ruiz}}]{Dan:2011}
{Dan}, M., {Rosswog}, S., {Guillochon}, J., \& {Ramirez-Ruiz}, E. 2011, \apj,
  737, 89, 1101.5132

\bibitem[{Fairhurst(2009)}]{Fairhurst:2009}
Fairhurst, S. 2009, New Journal of Physics, 11, 123006

\bibitem[{{Farr} {et~al.}(2017){Farr}, {Stevenson}, {Miller}, {Mandel}, {Farr},
  \& {Vecchio}}]{Farr:2017}
{Farr}, W.~M., {Stevenson}, S., {Miller}, M.~C., {Mandel}, I., {Farr}, B., \&
  {Vecchio}, A. 2017, \nat, 548, 426, 1706.01385

\bibitem[{{Finn}(1996)}]{Finn:1996}
{Finn}, L.~S. 1996, \prd, 53, 2878, arXiv:gr-qc/9601048

\bibitem[{{Freire} {et~al.}(2017){Freire}, {Ridolfi}, {Kramer}, {Jordan},
  {Manchester}, {Torne}, {Sarkissian}, {Heinke}, {D'Amico}, {Camilo},
  {Lorimer}, \& {Lyne}}]{Freire:2017}
{Freire}, P.~C.~C. {et~al.} 2017, \mnras, 471, 857, 1706.04908

\bibitem[{{Gair} {et~al.}(2011){Gair}, {Mandel}, {Miller}, \&
  {Volonteri}}]{Gair:2009ETrev}
{Gair}, J.~R., {Mandel}, I., {Miller}, M.~C., \& {Volonteri}, M. 2011, General
  Relativity and Gravitation, 43, 485, 0907.5450

\bibitem[{{Gair} {et~al.}(2009){Gair}, {Mandel}, {Sesana}, \&
  {Vecchio}}]{Gair:2009ET}
{Gair}, J.~R., {Mandel}, I., {Sesana}, A., \& {Vecchio}, A. 2009, Classical and
  Quantum Gravity, 26, 204009, 0907.3292

\bibitem[{{Gilfanov} \& {Bogd{\'a}n}(2010)}]{GilfanovBogdan:2010}
{Gilfanov}, M., \& {Bogd{\'a}n}, {\'A}. 2010, \nat, 463, 924, 1002.3359

\bibitem[{{Gonz{\'a}lez Hern{\'a}ndez} {et~al.}(2012){Gonz{\'a}lez
  Hern{\'a}ndez}, {Ruiz-Lapuente}, {Tabernero}, {Montes}, {Canal},
  {M{\'e}ndez}, \& {Bedin}}]{GonzalezHernandez:2012}
{Gonz{\'a}lez Hern{\'a}ndez}, J.~I., {Ruiz-Lapuente}, P., {Tabernero}, H.~M.,
  {Montes}, D., {Canal}, R., {M{\'e}ndez}, J., \& {Bedin}, L.~R. 2012, \nat,
  489, 533, 1210.1948

\bibitem[{{Graham} {et~al.}(2013){Graham}, {Hogan}, {Kasevich}, \&
  {Rajendran}}]{Graham:2013}
{Graham}, P.~W., {Hogan}, J.~M., {Kasevich}, M.~A., \& {Rajendran}, S. 2013,
  Physical Review Letters, 110, 171102, 1206.0818

\bibitem[{{Grover} {et~al.}(2014){Grover}, {Fairhurst}, {Farr}, {Mandel},
  {Rodriguez}, {Sidery}, \& {Vecchio}}]{Grover:2013}
{Grover}, K., {Fairhurst}, S., {Farr}, B.~F., {Mandel}, I., {Rodriguez}, C.,
  {Sidery}, T., \& {Vecchio}, A. 2014, \prd, 89, 042004, 1310.7454

\bibitem[{{Han} \& {Podsiadlowski}(2004)}]{PodsiadlowskiHan:2004}
{Han}, Z., \& {Podsiadlowski}, P. 2004, \mnras, 350, 1301, astro-ph/0309618

\bibitem[{{Harms} {et~al.}(2013){Harms}, {Slagmolen}, {Adhikari}, {Miller},
  {Evans}, {Chen}, {M{\"u}ller}, \& {Ando}}]{Harms:2013}
{Harms}, J., {Slagmolen}, B.~J.~J., {Adhikari}, R.~X., {Miller}, M.~C.,
  {Evans}, M., {Chen}, Y., {M{\"u}ller}, H., \& {Ando}, M. 2013, \prd, 88,
  122003, 1308.2074

\bibitem[{{Haster} {et~al.}(2016{\natexlab{a}}){Haster}, {Antonini},
  {Kalogera}, \& {Mandel}}]{Haster:2016}
{Haster}, C.-J., {Antonini}, F., {Kalogera}, V., \& {Mandel}, I.
  2016{\natexlab{a}}, \apj, 832, 192, 1606.07097

\bibitem[{{Haster} {et~al.}(2016{\natexlab{b}}){Haster}, {Wang}, {Berry},
  {Stevenson}, {Veitch}, \& {Mandel}}]{Haster:2015IMRI}
{Haster}, C.-J., {Wang}, Z., {Berry}, C.~P.~L., {Stevenson}, S., {Veitch}, J.,
  \& {Mandel}, I. 2016{\natexlab{b}}, \mnras, 457, 4499, 1511.01431

\bibitem[{{Hayden} {et~al.}(2010){Hayden}, {Garnavich}, {Kasen}, {Dilday},
  {Frieman}, {Jha}, {Lampeitl}, {Nichol}, {Sako}, {Schneider}, {Smith},
  {Sollerman}, \& {Wheeler}}]{Hayden:2010}
{Hayden}, B.~T. {et~al.} 2010, \apj, 722, 1691, 1008.4797

\bibitem[{{Hobbs} {et~al.}(2010){Hobbs}, {Archibald}, {Arzoumanian}, {Backer},
  {Bailes}, {Bhat}, {Burgay}, {Burke-Spolaor}, {Champion}, {Cognard}, {Coles},
  {Cordes}, {Demorest}, {Desvignes}, {Ferdman}, {Finn}, {Freire}, {Gonzalez},
  {Hessels}, {Hotan}, {Janssen}, {Jenet}, {Jessner}, {Jordan}, {Kaspi},
  {Kramer}, {Kondratiev}, {Lazio}, {Lazaridis}, {Lee}, {Levin}, {Lommen},
  {Lorimer}, {Lynch}, {Lyne}, {Manchester}, {McLaughlin}, {Nice}, {Oslowski},
  {Pilia}, {Possenti}, {Purver}, {Ransom}, {Reynolds}, {Sanidas}, {Sarkissian},
  {Sesana}, {Shannon}, {Siemens}, {Stairs}, {Stappers}, {Stinebring},
  {Theureau}, {van Haasteren}, {van Straten}, {Verbiest}, {Yardley}, \&
  {You}}]{PTA}
{Hobbs}, G. {et~al.} 2010, Classical and Quantum Gravity, 27, 084013, 0911.5206

\bibitem[{{Howell}(2011)}]{Howell:2011}
{Howell}, D.~A. 2011, Nature Communications, 2, 350, 1011.0441

\bibitem[{{K{\i}z{\i}ltan} {et~al.}(2017){K{\i}z{\i}ltan}, {Baumgardt}, \&
  {Loeb}}]{Kiziltan:2017}
{K{\i}z{\i}ltan}, B., {Baumgardt}, H., \& {Loeb}, A. 2017, \nat, 542, 203,
  1702.02149

\bibitem[{{Kopparapu} {et~al.}(2008){Kopparapu}, {Hanna}, {Kalogera},
  {O'Shaughnessy}, {Gonzalez}, {Brady}, \& {Fairhurst}}]{LIGOS3S4Galaxies}
{Kopparapu}, R.~K., {Hanna}, C.~R., {Kalogera}, V., {O'Shaughnessy}, R.,
  {Gonzalez}, G., {Brady}, P.~R., \& {Fairhurst}, S. 2008, \apj, 675, 1459

\bibitem[{{Kormendy} \& {Ho}(2013)}]{KormendyHo:2013}
{Kormendy}, J., \& {Ho}, L.~C. 2013, \araa, 51, 511, 1304.7762

\bibitem[{{Kormendy} \& {Richstone}(1995)}]{Kormendy:1995}
{Kormendy}, J., \& {Richstone}, D. 1995, \araa, 33, 581

\bibitem[{{Kozai}(1962)}]{Kozai:1962}
{Kozai}, Y. 1962, \aj, 67, 591

\bibitem[{{Kushnir} {et~al.}(2013){Kushnir}, {Katz}, {Dong}, {Livne}, \&
  {Fern{\'a}ndez}}]{Kushnir:2013}
{Kushnir}, D., {Katz}, B., {Dong}, S., {Livne}, E., \& {Fern{\'a}ndez}, R.
  2013, \apjl, 778, L37, 1303.1180

\bibitem[{{Lasky} {et~al.}(2016){Lasky}, {Mingarelli}, {Smith}, {Giblin},
  {Thrane}, {Reardon}, {Caldwell}, {Bailes}, {Bhat}, {Burke-Spolaor}, {Dai},
  {Dempsey}, {Hobbs}, {Kerr}, {Levin}, {Manchester}, {Os{\l}owski}, {Ravi},
  {Rosado}, {Shannon}, {Spiewak}, {van Straten}, {Toomey}, {Wang}, {Wen},
  {You}, \& {Zhu}}]{Lasky:2016}
{Lasky}, P.~D. {et~al.} 2016, Physical Review X, 6, 011035, 1511.05994

\bibitem[{{Lidov}(1962)}]{Lidov:1962}
{Lidov}, M.~L. 1962, Planet. Space Sci., 9, 719

\bibitem[{{Littenberg} {et~al.}(2015){Littenberg}, {Farr}, {Coughlin},
  {Kalogera}, \& {Holz}}]{Littenberg:2015}
{Littenberg}, T.~B., {Farr}, B., {Coughlin}, S., {Kalogera}, V., \& {Holz},
  D.~E. 2015, \apjl, 807, L24, 1503.03179

\bibitem[{{Livio}(2000)}]{Livio:2000}
{Livio}, M. 2000, in Type Ia Supernovae, Theory and Cosmology, ed. J.~C.
  {Niemeyer} \& J.~W. {Truran}, 33, astro-ph/9903264

\bibitem[{{Luo} {et~al.}(2016){Luo}, {Chen}, {Duan}, {Gong}, {Hu}, {Ji}, {Liu},
  {Mei}, {Milyukov}, {Sazhin}, {Shao}, {Toth}, {Tu}, {Wang}, {Wang}, {Yeh},
  {Zhan}, {Zhang}, {Zharov}, \& {Zhou}}]{TianQin}
{Luo}, J. {et~al.} 2016, Classical and Quantum Gravity, 33, 035010, 1512.02076

\bibitem[{{Magorrian} {et~al.}(1998){Magorrian}, {Tremaine}, {Richstone},
  {Bender}, {Bower}, {Dressler}, {Faber}, {Gebhardt}, {Green}, {Grillmair},
  {Kormendy}, \& {Lauer}}]{Magorrian:1998}
{Magorrian}, J. {et~al.} 1998, \aj, 115, 2285, astro-ph/9708072

\bibitem[{{Mandel} {et~al.}(2008){Mandel}, {Brown}, {Gair}, \&
  {Miller}}]{Mandel:2008}
{Mandel}, I., {Brown}, D.~A., {Gair}, J.~R., \& {Miller}, M.~C. 2008, \apj,
  681, 1431, 0705.0285

\bibitem[{{Mandel} {et~al.}(2015){Mandel}, {Haster}, {Dominik}, \&
  {Belczynski}}]{Mandel:2015}
{Mandel}, I., {Haster}, C.-J., {Dominik}, M., \& {Belczynski}, K. 2015, \mnras,
  450, L85, 1503.03172

\bibitem[{{Mandic} {et~al.}(2012){Mandic}, {Thrane}, {Giampanis}, \&
  {Regimbau}}]{Mandic:2012}
{Mandic}, V., {Thrane}, E., {Giampanis}, S., \& {Regimbau}, T. 2012, Physical
  Review Letters, 109, 171102, 1209.3847

\bibitem[{{Mennekens} {et~al.}(2010){Mennekens}, {Vanbeveren}, {De Greve}, \&
  {De Donder}}]{Mennekens:2010}
{Mennekens}, N., {Vanbeveren}, D., {De Greve}, J.~P., \& {De Donder}, E. 2010,
  \aap, 515, A89, 1003.2491

\bibitem[{{Miller} \& {Colbert}(2004)}]{MillerColbert:2004}
{Miller}, M.~C., \& {Colbert}, E.~J.~M. 2004, International Journal of Modern
  Physics D, 13, 1, arXiv:astro-ph/0308402

\bibitem[{{Nielsen} {et~al.}(2014){Nielsen}, {Nelemans}, {Voss}, \&
  {Toonen}}]{Nielsen:2014}
{Nielsen}, M.~T.~B., {Nelemans}, G., {Voss}, R., \& {Toonen}, S. 2014, \aap,
  563, A16, 1310.2170

\bibitem[{{Nugent} {et~al.}(2011){Nugent}, {Sullivan}, {Cenko}, {Thomas},
  {Kasen}, {Howell}, {Bersier}, {Bloom}, {Kulkarni}, {Kandrashoff},
  {Filippenko}, {Silverman}, {Marcy}, {Howard}, {Isaacson}, {Maguire},
  {Suzuki}, {Tarlton}, {Pan}, {Bildsten}, {Fulton}, {Parrent}, {Sand},
  {Podsiadlowski}, {Bianco}, {Dilday}, {Graham}, {Lyman}, {James}, {Kasliwal},
  {Law}, {Quimby}, {Hook}, {Walker}, {Mazzali}, {Pian}, {Ofek}, {Gal-Yam}, \&
  {Poznanski}}]{Nugent:2011}
{Nugent}, P.~E. {et~al.} 2011, \nat, 480, 344, 1110.6201

\bibitem[{{Pakmor} {et~al.}(2012){Pakmor}, {Kromer}, {Taubenberger}, {Sim},
  {R{\"o}pke}, \& {Hillebrandt}}]{Pakmor:2012}
{Pakmor}, R., {Kromer}, M., {Taubenberger}, S., {Sim}, S.~A., {R{\"o}pke},
  F.~K., \& {Hillebrandt}, W. 2012, \apjl, 747, L10, 1201.5123

\bibitem[{{Pasham} {et~al.}(2014){Pasham}, {Strohmayer}, \&
  {Mushotzky}}]{Pasham:2014}
{Pasham}, D.~R., {Strohmayer}, T.~E., \& {Mushotzky}, R.~F. 2014, \nat, 513, 74

\bibitem[{{Peters}(1964)}]{Peters:1964}
{Peters}, P.~C. 1964, Physical Review, 136, 1224

\bibitem[{{Poisson} \& {Will}(1995)}]{PoissonWill:1995}
{Poisson}, E., \& {Will}, C.~M. 1995, \prd, 52, 848, arXiv:gr-qc/9502040

\bibitem[{{Punturo} {et~al.}(2010){Punturo}, {Abernathy}, {Acernese}, {Allen},
  {Andersson}, {Arun}, {Barone}, {Barr}, {et~al.}}]{ET}
{Punturo}, M. {et~al.} 2010, Classical and Quantum Gravity, 27, 084007

\bibitem[{{Rodriguez} {et~al.}(2012){Rodriguez}, {Mandel}, \&
  {Gair}}]{Rodriguez:2012}
{Rodriguez}, C.~L., {Mandel}, I., \& {Gair}, J.~R. 2012, \prd, 85, 062002,
  1112.1404

\bibitem[{{Sathyaprakash} {et~al.}(2012){Sathyaprakash}, {Abernathy},
  {Acernese}, {Ajith}, {Allen}, {Amaro-Seoane}, {Andersson}, {Aoudia}, {Arun},
  {Astone}, \& et~al.}]{ET:2012}
{Sathyaprakash}, B. {et~al.} 2012, Classical and Quantum Gravity, 29, 124013,
  1206.0331

\bibitem[{{Sesana} {et~al.}(2011){Sesana}, {Gair}, {Berti}, \&
  {Volonteri}}]{Sesana:2011}
{Sesana}, A., {Gair}, J., {Berti}, E., \& {Volonteri}, M. 2011, \prd, 83,
  044036, 1011.5893

\bibitem[{{Sesana} {et~al.}(2009){Sesana}, {Gair}, {Mandel}, \&
  {Vecchio}}]{Sesana:2009ET}
{Sesana}, A., {Gair}, J., {Mandel}, I., \& {Vecchio}, A. 2009, \apjl, 698,
  L129, 0903.4177

\bibitem[{{Sesana} {et~al.}(2008){Sesana}, {Vecchio}, {Eracleous}, \&
  {Sigurdsson}}]{Sesana:2008}
{Sesana}, A., {Vecchio}, A., {Eracleous}, M., \& {Sigurdsson}, S. 2008, ArXiv
  e-prints, 0806.0624, 0806.0624

\bibitem[{{Stevenson} {et~al.}(2017){Stevenson}, {Berry}, \&
  {Mandel}}]{Stevenson:2017spin}
{Stevenson}, S., {Berry}, C.~P.~L., \& {Mandel}, I. 2017, \mnras, 471, 2801,
  1703.06873

\bibitem[{{Stone} {et~al.}(2017){Stone}, {Metzger}, \& {Haiman}}]{Stone:2016}
{Stone}, N.~C., {Metzger}, B.~D., \& {Haiman}, Z. 2017, \mnras, 464, 946,
  1602.04226

\bibitem[{{Takahashi} \& {Nakamura}(2003)}]{TakahashiNakamura:2003}
{Takahashi}, R., \& {Nakamura}, T. 2003, \apjl, 596, L231, astro-ph/0307390

\bibitem[{{The LIGO Scientific Collaboration} \& {The Virgo
  Collaboration}(2017)}]{GW170817}
{The LIGO Scientific Collaboration}, \& {The Virgo Collaboration}. 2017, ArXiv
  e-prints, 1710.05832

\bibitem[{{Verbiest} {et~al.}(2016){Verbiest}, {Lentati}, {Hobbs}, {van
  Haasteren}, {Demorest}, {Janssen}, {Wang}, {Desvignes}, {Caballero}, {Keith},
  {Champion}, {Arzoumanian}, {Babak}, {Bassa}, {Bhat}, {Brazier}, {Brem},
  {Burgay}, {Burke-Spolaor}, {Chamberlin}, {Chatterjee}, {Christy}, {Cognard},
  {Cordes}, {Dai}, {Dolch}, {Ellis}, {Ferdman}, {Fonseca}, {Gair},
  {Garver-Daniels}, {Gentile}, {Gonzalez}, {Graikou}, {Guillemot}, {Hessels},
  {Jones}, {Karuppusamy}, {Kerr}, {Kramer}, {Lam}, {Lasky}, {Lassus},
  {Lazarus}, {Lazio}, {Lee}, {Levin}, {Liu}, {Lynch}, {Lyne}, {Mckee},
  {McLaughlin}, {McWilliams}, {Madison}, {Manchester}, {Mingarelli}, {Nice},
  {Os{\l}owski}, {Palliyaguru}, {Pennucci}, {Perera}, {Perrodin}, {Possenti},
  {Petiteau}, {Ransom}, {Reardon}, {Rosado}, {Sanidas}, {Sesana}, {Shaifullah},
  {Shannon}, {Siemens}, {Simon}, {Smits}, {Spiewak}, {Stairs}, {Stappers},
  {Stinebring}, {Stovall}, {Swiggum}, {Taylor}, {Theureau}, {Tiburzi},
  {Toomey}, {Vallisneri}, {van Straten}, {Vecchio}, {Wang}, {Wen}, {You},
  {Zhu}, \& {Zhu}}]{IPTA}
{Verbiest}, J.~P.~W. {et~al.} 2016, \mnras, 458, 1267, 1602.03640

\bibitem[{{Volonteri}(2010)}]{Volonteri:2010}
{Volonteri}, M. 2010, The Astronomy and Astrophysics Review, 18, 279, 1003.4404

\bibitem[{{Volonteri} {et~al.}(2003){Volonteri}, {Haardt}, \& {Madau}}]{VHM}
{Volonteri}, M., {Haardt}, F., \& {Madau}, P. 2003, \apj, 582, 559

\bibitem[{{Yoon} {et~al.}(2007){Yoon}, {Podsiadlowski}, \&
  {Rosswog}}]{Yoon:2007}
{Yoon}, S.-C., {Podsiadlowski}, P., \& {Rosswog}, S. 2007, \mnras, 380, 933,
  0704.0297

\bibitem[{{Zevin} {et~al.}(2017){Zevin}, {Pankow}, {Rodriguez}, {Sampson},
  {Chase}, {Kalogera}, \& {Rasio}}]{Zevin:2017}
{Zevin}, M., {Pankow}, C., {Rodriguez}, C.~L., {Sampson}, L., {Chase}, E.,
  {Kalogera}, V., \& {Rasio}, F.~A. 2017, \apj, 846, 82, 1704.07379

\end{thebibliography}

\end{document}